\documentclass[a4paper,10pt,twoside]{cpc-hepnp}

\usepackage{multicol}
\usepackage{graphicx}
\usepackage{booktabs}
\usepackage{amssymb,bm,mathrsfs,bbm,amscd}
\usepackage[tbtags]{amsmath}
\usepackage{lastpage}
\usepackage{CJK}
\usepackage{subcaption}
\usepackage{hepunits}
\usepackage{hyperref}
\usepackage{lineno}

\begin{document}
\begin{CJK*}{UTF8}{gbsn}


\fancyhead[c]{\small Submitted to Chinese Physics C} 
\fancyfoot[C]{\small \thepage}
\footnotetext[0]{ }

\title{Design and optimization of resistive anode for a two-dimensional imaging triple-GEM detector\thanks{Supported by National Natural Science
Foundation of China (11375219) }}

\author{%
      JU Xu-Dong(鞠旭东)$^{1,2,3;1)}$\email{juxd@ihep.ac.cn}%
\quad DONG Ming-Yi(董明义)$^{1,2;2)}$\email{dongmy@ihep.ac.cn}%
\quad ZHAO Yi-Chen(赵逸琛)$^{4}$\\
\quad ZHOU Chuan-Xing(周传兴)$^{1,2,3}$
\quad OUYANG Qun(欧阳群)$^{1,2}$
}
\maketitle

\address{%
$^1$ State Key Laboratory of Particle Detection and Electronics, Beijing 100049, China\\
$^2$ Institute of High Energy Physics, Chinese Academy of Sciences, Beijing 100049, China\\
$^3$ University of Chinese Academy of Sciences, Beijing 100049, China\\
$^4$ Zhengzhou University, Zhengzhou 450001, China\\
}

\begin{abstract}
The optimization of resistive anode for two dimensional imaging detectors which consists of a series of high resistive square pads surrounding by low resistive strips has been studied by both numerical simulations and experimental tests. It has been found that to obtain good detector performance, the resistance ratio of the pad to the strip should be larger than 5, the nonuniformity of the pad surface resistivity had better be less than $20\%$, a smaller pad width leads to a smaller spatial resolution and when the pad width is $6mm$, the spatial resolution ($\sigma$) can reach about $105{\mu}m$. Based on the study results, a 2-D GEM detector prototype with the optimized resistive anode is constructed and a good imaging performance is achieved.
\end{abstract}



\footnotetext[0]{\hspace*{-3mm}\raisebox{0.3ex}{$\scriptstyle\copyright$}2013
Chinese Physical Society and the Institute of High Energy Physics
of the Chinese Academy of Sciences and the Institute
of Modern Physics of the Chinese Academy of Sciences and IOP Publishing Ltd}%

\begin{multicols}{2}

\section{Introduction}
Gas Electron Multiplier (GEM)\cite{ref-1}, an important member of new Micro-Pattern Gaseous Detectors (MPGD), has been widely used in particle physical experiments such as CERN-COMPASS\cite{ref-2} because of its premium properties like good spatial resolution, highly counting rate, low material budget and so on. Compared to traditional gaseous detectors like drift chambers, the GEM detector has an advantage in the design of its readout electrode which separates from the multiplication region, therefore various readout methods can be used for the GEM detector such as 1-D strip, 2-D strip and pixel readout methods\cite{ref-3}. However general readout methods lead to a large number of electronics channels, for example there are about 30,000 channels in the GEM detector of KLOE2\cite{ref-4}. And thus, lots of new readout structures have been developed to reduce the electronics.
\par\indent
According to the four-corner resistive readout method\cite{ref-5,ref-6}, which is widely used in Position Sensitive silicon Detectors (PSD)\cite{ref-7}, Sarvestani et al.\cite{ref-8} developed a 2-D interpolating resistive readout structure for Micro-CAT. This readout structure can provide good spatial resolution with an enormous reduction of electronics channels, especially compared to the pixel readout. Based on this concept, we developed the triple-GEM detector with 2-D resistive anode\cite{ref-9}. It has been found that the resistive anode has a strong effect on the detector performance, especially, some resistive anodes lead to serious imaging distortions. To study the effect of the resistive anode, a series of anodes with different parameters has been fabricated and tested. In this paper, the optimization of resistive anode will be studied by both simulations and experiments.
\section{Detector setup and simulation model}
\subsection{Detector setup and principles}
The experimental setup is similar to what has been presented in Ref.\cite{ref-9} except that the DAQ system develops from CAMAC-Bus to VME-Bus. Fig.~\ref{fig_Electronics} shows the schematic of the detector setup. Three cascading standard GEM foils from CERN are working as gas gain device. Electrons generated in the drift region by incoming particles are multiplied in the holes of the GEM foils and then drifted to the resistive anode. Charges induced by electron clouds in the induction region will diffuse on the resistive anode and be collected by the readout nodes at the cell corners. For the detector prototype study, 16 readout nodes covering $3\times3$ cells are equipped with charge sensitive amplifiers and 12-bit Peak ADCs which are linked to a PC through VME-Bus.
\begin{center}
\includegraphics[width=8cm]{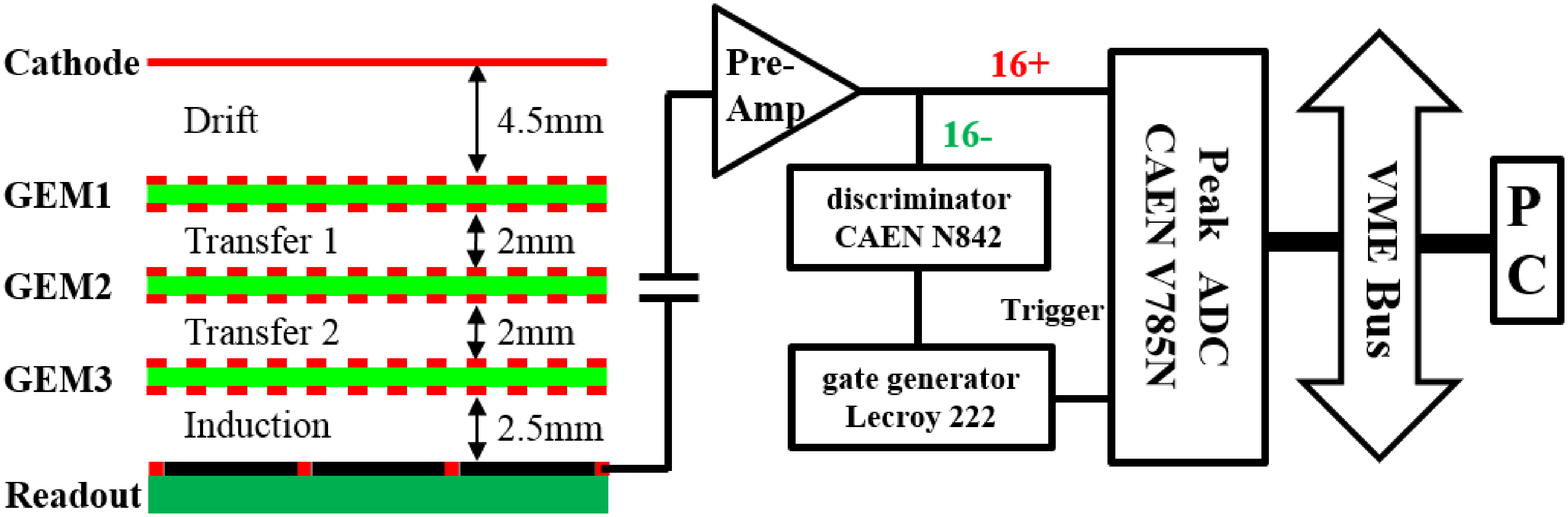}
\figcaption{\label{fig_Electronics}   Schematic of detector setup }
\end{center}
\par\indent
The resistive anode is manufactured by the thick-film resistor process on ceramic board and the structure is shown in Fig.~\ref{fig_Ceramic}. The sensitive area is divided into squared cells. Each cell consists of three parts: the high surface resistivity pad in the cell center with typical value of $100\sim1000k\Omega/\square$ and $7.8mm\times7.8mm$, low surface resistivity strips surrounding the squared pad with typical value of $1\sim10k\Omega/\square$ and $8mm\times0.2mm$, and readout nodes at the four corners of the cell.
\begin{center}
\includegraphics[width=8cm]{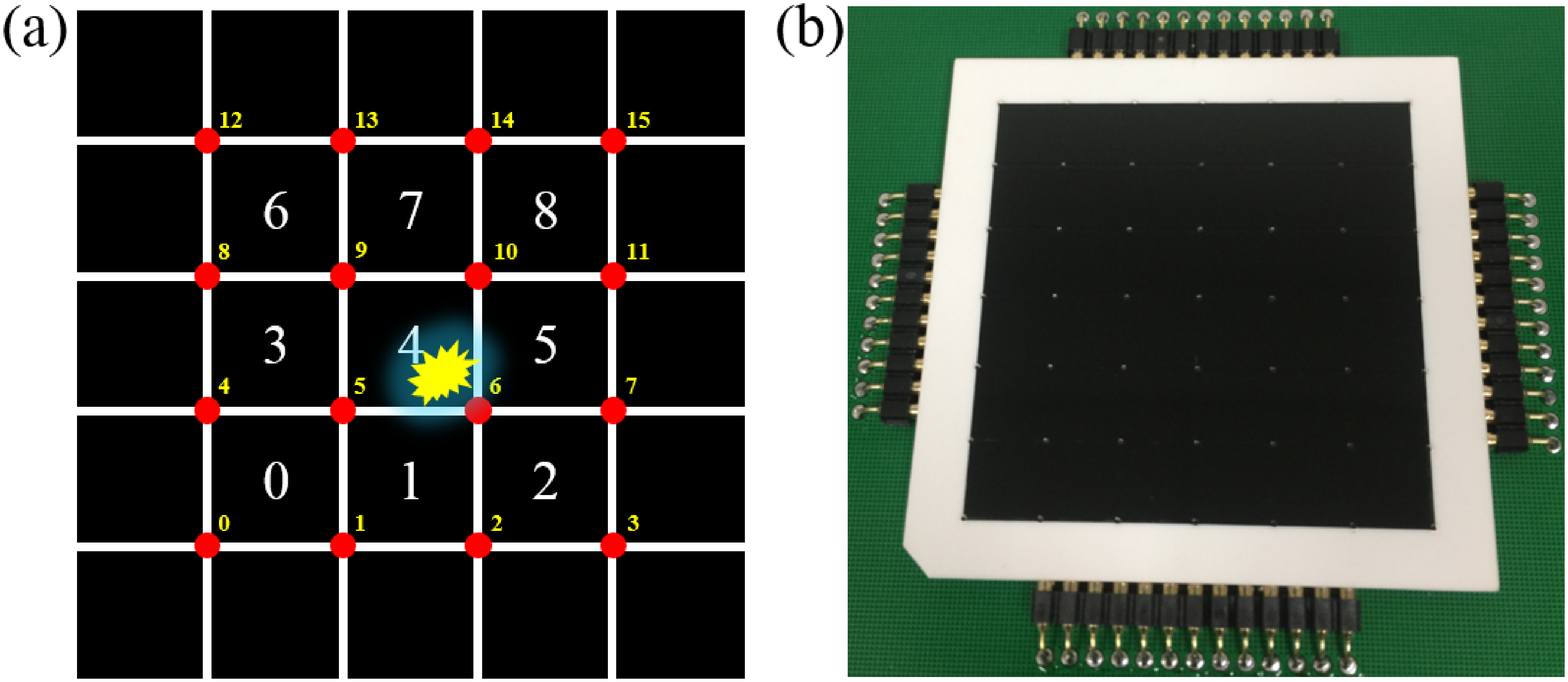}
\figcaption{\label{fig_Ceramic}  (a)schematic structure of resistive anode consisting of squared pads with high surface resistivity which are surrounded by  strips with low surface resistivity; (b)photo of the resistive ceramic board including $6\times6$ cells which is used in the experiments.}
\end{center}
\par\indent
It's the property of the resistive anode that has a determinate influence on the detector performance. Considering the structure of the resistive anode in Fig.~\ref{fig_Ceramic}, parameters that may affect the quality of resistive anode are:
\begin{itemize}
\item[-] Resistance ratio of the pad to the strip
\item[-] Dimensions of the pad and the strip
\item[-] Uniformity of the pad surface resistivity
\item[-] \ldots
\end{itemize}
A series of resistive anode boards with different parameters has been fabricated to study the influence of above factors on the detector performance, especially on the spatial resolution.
\subsection{Simulation model}
In order to understand the mechanics of charge diffusion on the resistive anode as well as to provide basis for the optimization of the anode structure, a numerical simulation model has been developed\cite{ref-10,ref-11}. By using the differential Ohm's law,
\begin{equation}
  \vec{j}(x,y,t)=-\sigma(x,y)\nabla V(x,y,t)
\end{equation}
where $\vec{j}(x,y,t)$ represents the wire current density, $\sigma(x,y)$ represents the surface conductivity and $V(x,y,t)$ represents the potential, and the current conservation law,
\begin{equation}
  \nabla\cdot\vec{j}(x,y,t)+c\frac{\partial V(x,y,t)}{\partial t}=0
\end{equation}
the diffusion equation can be obtained as
\begin{equation}\label{eq:diffusion}
    c\frac{\partial V(x,y,t)}{\partial t}-\nabla\cdot(\sigma(x,y)\nabla V(x,y,t))=I(x,y,t)
\end{equation}
where $I(x,y,t)$ represents the detector signal induced by electron clouds in the induction region. A simulation using Garfield++\cite{ref-12} is developed to evaluate $I(x,y,t)$ which is treated as a space and time dependent driving source
\begin{equation}\label{eq:I}
    I(x,y,t)=I_0S(x,y)T(t)
\end{equation}
and the result is shown in Fig.~\ref{fig_Garfield}. Therefore, in Eq.~({\ref{eq:I}}), the spatial part of the detector signal ($S(x,y)$) is simplified as a gaussian distribution with $\sigma=0.25mm$ and the time part ($T(t)$) is simplified as a uniform distribution.
\par\indent
The charge diffusion on the resistive anode is then abstracted as the solving of a partial differential equation with the constraint condition that voltage at all readout nodes equals zero.  The explicit finite difference method (FDM) is used to solve the equation in this work\cite{ref-10,ref-11}.
\begin{center}
\includegraphics[width=8cm]{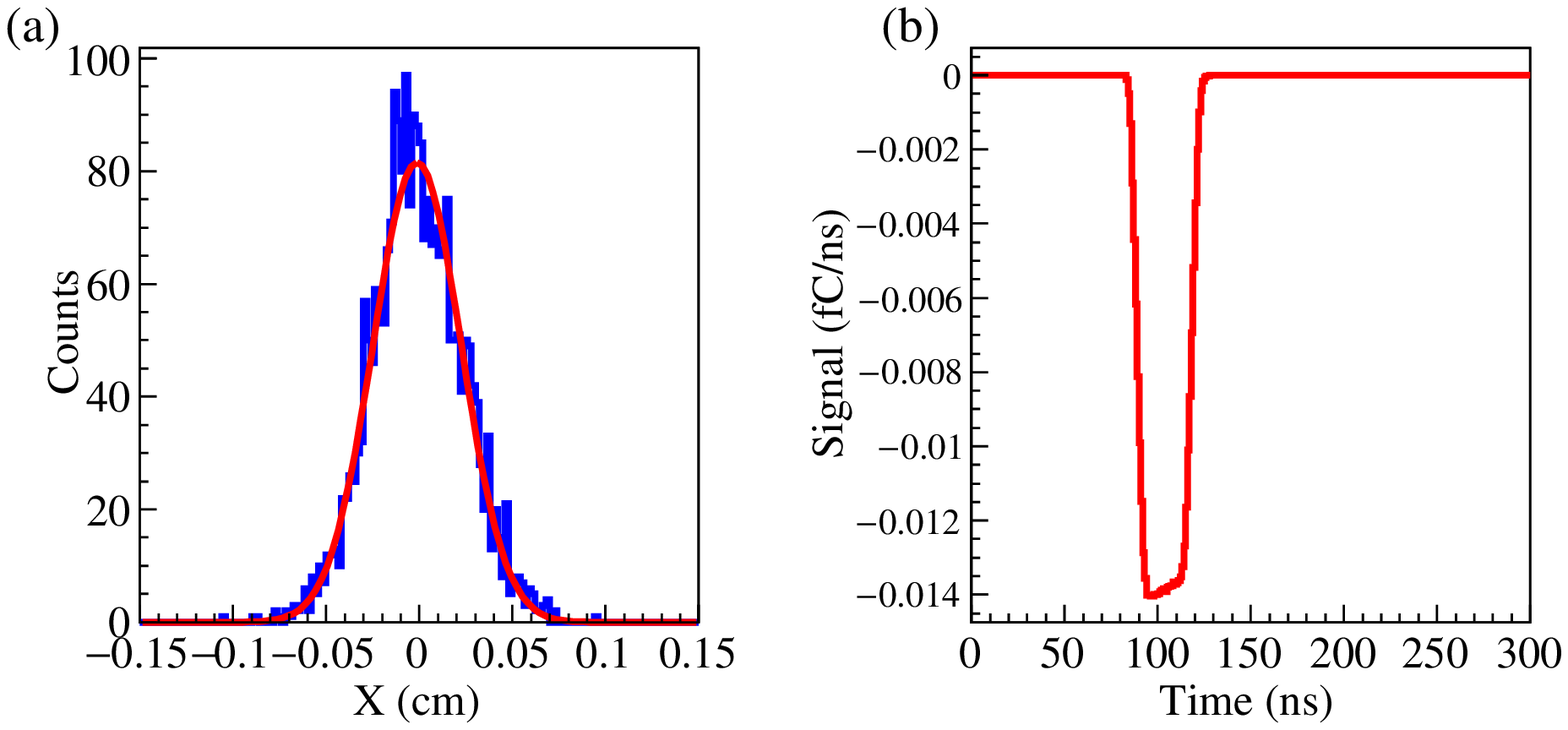}
\figcaption{\label{fig_Garfield}   Garfield++ simulation results of the detector signal. (a)spatial distribution of electrons on the anode in the x-direction; (b)time distribution of charges collected by the anode}
\end{center}
\subsection{Reconstruction method}
For PSDs, the typical reconstruction method is the charge center of gravity method (COG). Considering the structural characteristics of the 2-D resistive anode readout method, the general reconstruction method is the so-called 4-node-method\cite{ref-7}. It uses the 4 nodes of the fired pad to reconstruct the hit position like
\begin{eqnarray}\label{eq:Reconstruction}
    X_4=\frac{\sum_{i=1}^{4}x_i(Q_i-Q_{0i})}{\sum_{i=1}^{4}(Q_i-Q_{0i})}\\
    Y_4=\frac{\sum_{i=1}^{4}y_i(Q_i-Q_{0i})}{\sum_{i=1}^{4}(Q_i-Q_{0i})}
\end{eqnarray}
where $x_i$ and $y_i$ represents the position of readout nodes, $Q_i$ represents charges obtained by electronics system and $Q_0i$ represents the baseline of the corresponding readout channel.
\par\indent
To reduce the distortion caused by charge loss from the fired pad to the adjacent pads, some improved methods have been developed\cite{ref-11,ref-13}. These relatively complex reconstruction methods can lead to good reconstruction results by using more information from the nodes on the adjacent pads.
\section{Experiment and simulation results}
\subsection{Resistance ratio of pad to strip}
In the experiments, it's found that the resistance ratio of the pad to the strip seriously affects the detector imaging performance. This ratio is defined as $R_{{\square}P}/(R_{{\square}L}{\cdot}N)$, where $R_{{\square}P}$ and $R_{{\square}L}$ is the surface resistivity of the pad and the strip respectively, $N$ is the aspect ratio of the strip.
\end{multicols}
\ruleup
\begin{center}
\includegraphics[width=16cm]{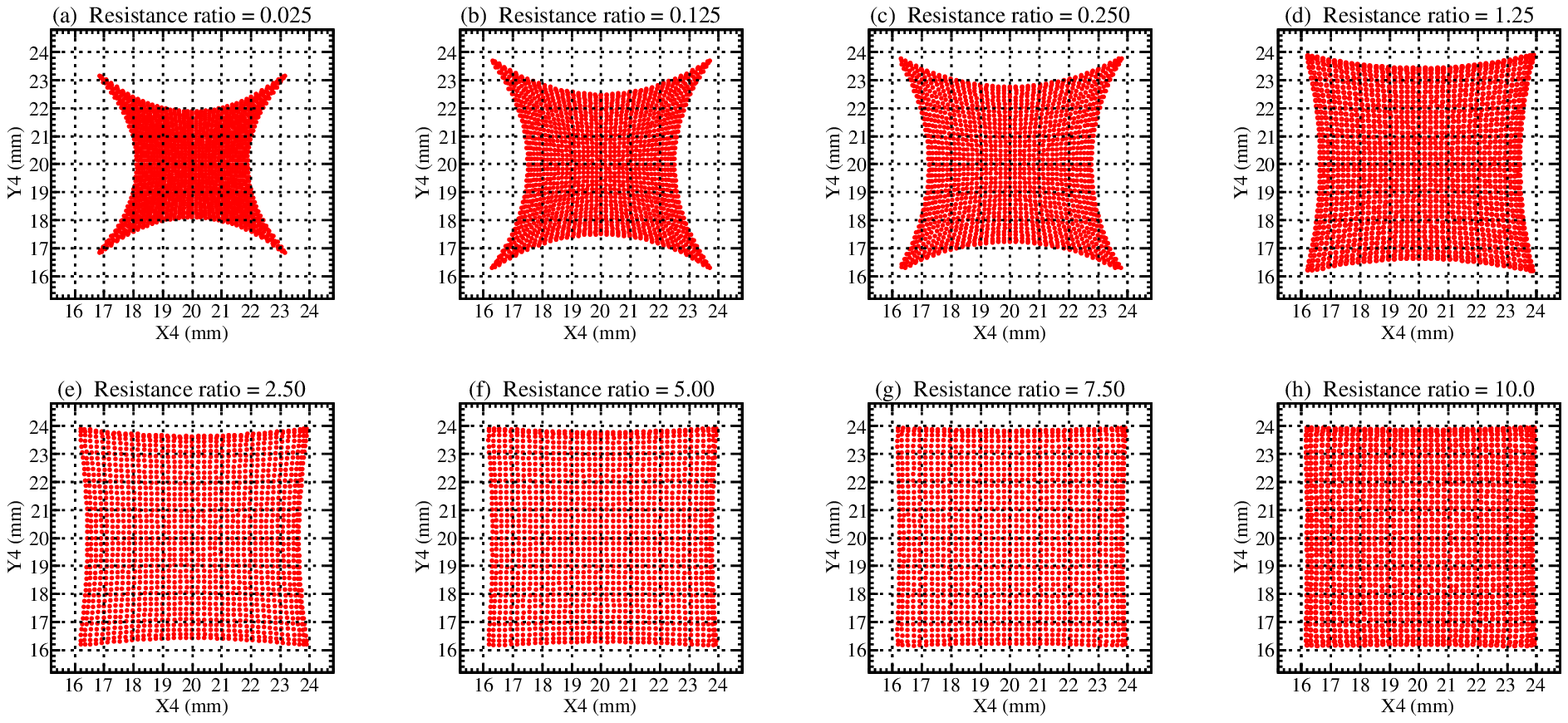}
\figcaption{\label{fig_ResistanceRatio} Simulation results of different $R_{{\square}P}/(R_{{\square}L}{\cdot}N)$. The size of the pad and the strip is $8mm\times8mm$ and $8mm\times0.2mm$, respectively. $R_{{\square}L}$ is fixed to be $1k\Omega/\square$, and $R_{{\square}P}$ changes from $1k\Omega/\square$ to $400k\Omega/\square$ from (a) to (h).}
\end{center}
\ruledown
\begin{multicols}{2}
\par\indent
Fig.~\ref{fig_ResistanceRatio} shows the simulation results of different $R_{{\square}P}/(R_{{\square}L}{\cdot}N)$. There is pincushion distortion in the imaging results, especially near the edge of the pad, the distortion is more serious. The main reason is that part of the charges will diffuse across the strip boundary and be collected by neighbouring pads. There're about 40\% charge loss (percent of charges collected by non-fired pads) on the edge of the pad when $R_{{\square}P}=R_{{\square}L}$(Fig.~\ref{fig_ResistanceRatio}-(a)). 
\par\indent
On the other hand, from Fig.~\ref{fig_ResistanceRatio}, we can also see that the integral distortion becomes less serious with a larger $R_{{\square}P}/(R_{{\square}L}{\cdot}N)$. In other words, a larger deviation between the resistance of the pad and the strip can help to improve the detector imaging performance, and when $R_{{\square}P}/(R_{{\square}L}{\cdot}N)>5$, this improvement slows down.
\par\indent
Based on the simulation results, some resistive anodes with different $R_{{\square}P}/(R_{{\square}L}{\cdot}N)$ have been fabricated and tested in a GEM detector. The detector imaging performance is shown in Fig.~\ref{fig_PadCompare}. Compared to the image obtained by using anode with $R_{{\square}P}/(R_{{\square}L}{\cdot}N)\thickapprox3$, the pincushion distortion is smaller for that with $R_{{\square}P}/(R_{{\square}L}{\cdot}N)\thickapprox5$, and this is consistant to the simulation results. 
\par\indent
In addition to the influence on the imaging performance, the distortion affects the spatial resolution as well. Fig.~\ref{fig_SRScan} shows the scan of the spatial resolution ($\sigma$, See in Sec.~\ref{Sec-Sigma}) of the detector equipped with the corresponding resistive boards in Fig.~\ref{fig_PadCompare}. The spatial resolution decreases from the edge to the middle of one single pad because of the distortion. For a larger $R_{{\square}P}/(R_{{\square}L}{\cdot}N)$, $\sigma$ is smaller and also much more uniform. 
\par\indent
Furthermore, considering the charge diffusion time on the resistive anode, especially for anodes with high surface resistivity\cite{ref-10}, a typical range for the surface resistivity of the Pad is $100\sim1000k\Omega/\square$ and the recommended value of $R_{{\square}P}/(R_{{\square}L}{\cdot}N)$ is larger than 5.
\begin{center}
\includegraphics[width=8cm]{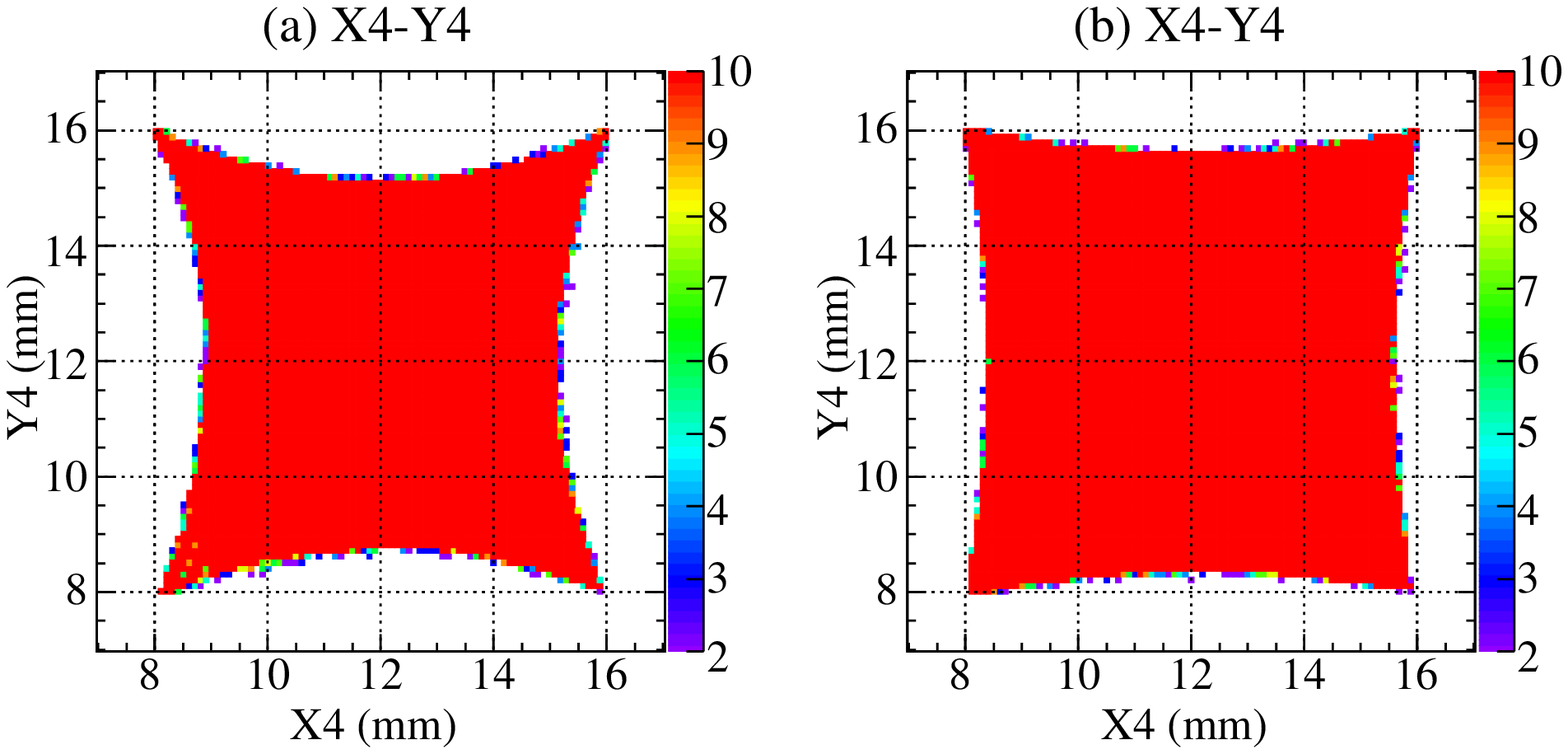}
\figcaption{\label{fig_PadCompare} The imaging of detector using resistive anodes with different $R_{{\square}P}/(R_{{\square}L}{\cdot}N)$. The size of the pad and the strip is $8mm\times8mm$ and $8mm\times0.2mm$, respectively. (a)$R_{{\square}P}/(R_{{\square}L}{\cdot}N)\thickapprox3$; (b)$R_{{\square}P}/(R_{{\square}L}{\cdot}N)\thickapprox5$.}
\end{center}
 \begin{center}
\includegraphics[width=8cm]{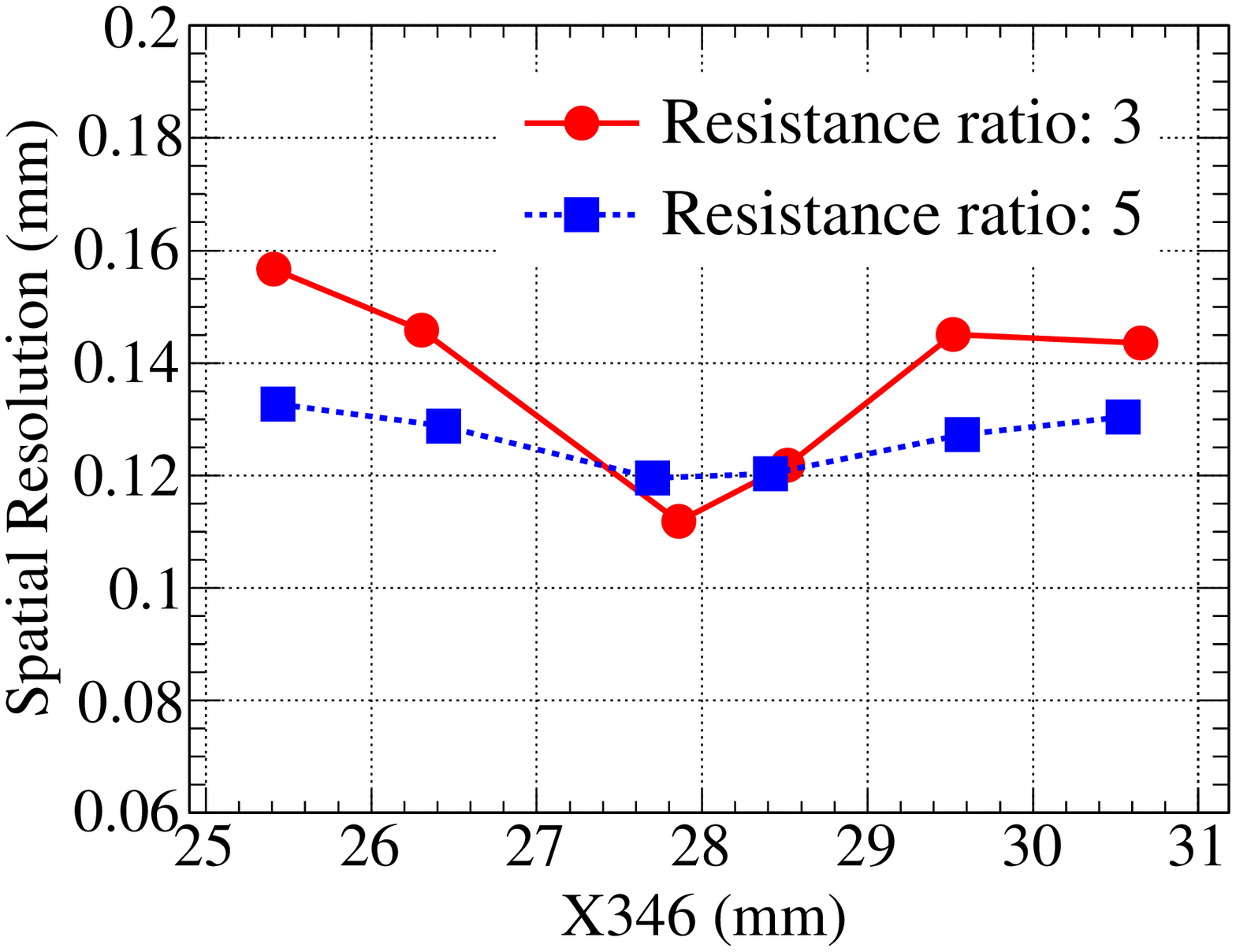}
\figcaption{\label{fig_SRScan} Spatial resolution scan of one single pad by 1mm step for the resistive anodes in Fig.~\ref{fig_PadCompare}. Results are getting from a $40{\mu}m$-slot by using X-ray machine.}
\end{center}
\subsection{Dimensions of pad}
\label{Sec-Sigma}
For the strip and the pixel readout methods, the pitch between strips or pixels has a direct effect on the spatial resolution. Accordingly, to study the effect of the pad width on the spatial resolution, resistive anodes with pad dimension of $6mm\times6mm$, $8mm\times8mm$ and $10mm\times10mm$ are fabricated and tested. The measurement is using X-ray machine with a $40{\mu}m$-slot. A composite 2-gaussian model is used to fit the counts distribution as shown in Fig.~\ref{fig_SR} and the spatial resolution is calculated as
\begin{equation}
  \sigma=\sqrt{g\cdot\sigma_S^2 + (1-g)\cdot\sigma_B^2}
\end{equation}
where $\sigma$ is the composite spatial resolution, $\sigma_S$ is the signal part, $\sigma_B$ is the background part, and $g$ is the fraction of the signal\cite{ref-14}.
\par\indent
Tbl.~\ref{tab_SR} shows the spatial resolution of the detector using resistive anode with different pad width. From Tbl.~\ref{tab_SR}, it can be seen that a better spatial resolution can be obtained by using a smaller pad width. However, for the detector with the same sized resisitive anode, a smaller pad width also means a little more readout channels. And thus the selection of the pad width should be a compromise between the spatial resolution and the electronics channels according to the actual application requirements.
\begin{center}
\includegraphics[width=8cm]{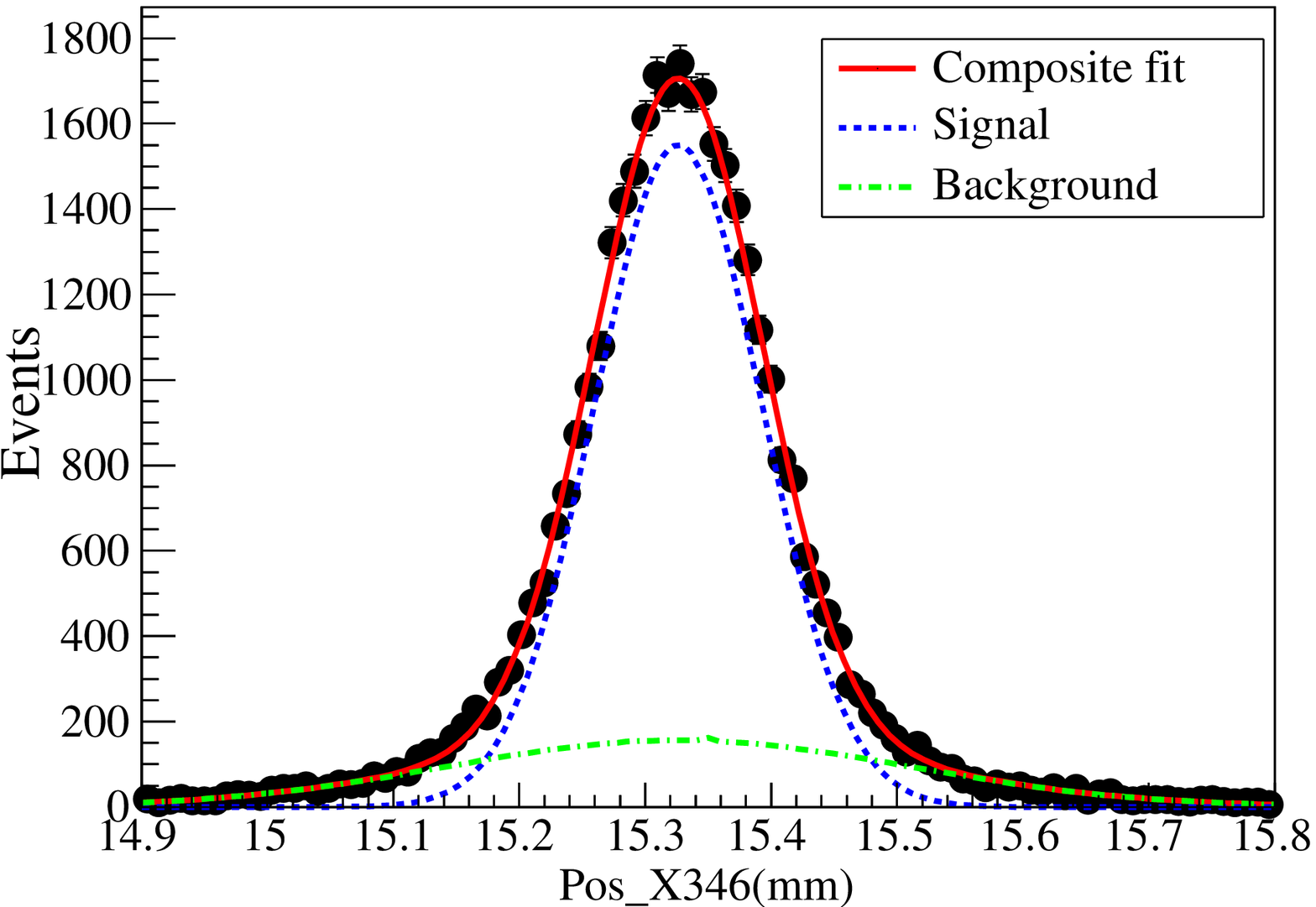}
\figcaption{\label{fig_SR} Spatial resolution of the detector using resistive anode with $6mm\times6mm$ sized pads. Results are getting from a $40{\mu}m$-slot by using X-ray machine.}
\end{center}
\begin{center}
\tabcaption{ \label{tab_SR}  Spatial resolution with different pad width.}
\footnotesize
\begin{tabular*}{80mm}{c@{\extracolsep{\fill}}ccc}
\toprule Width(mm) & $\sigma({\mu}m)$  & $\sigma_S({\mu}m)$ \\
\hline
6\hphantom{0} & 103.4 & \hphantom{0}66.7  \\
8\hphantom{0} & 112.2 & \hphantom{0}80.0  \\
10            & 145.1 & 109.9  \\
\bottomrule
\end{tabular*}
\end{center}
\subsection{Nonuniformity of the pad surface resistivity}
Nonuniformity of the pad surface resistivity is another important factor that can impact on the detector performance. Fig.~\ref{fig_Nonuniformity} shows the influence of nonuniformity on the imaging performance by simulation. To introduce the nonuniformity effect into the simulation, one single pad is divided into four parts and a constant ratio of the original value of the pad surface resistivity is added to each part. For example, in Fig.~\ref{fig_Nonuniformity}-(a), +5\%, -5\%, +5\%, -5\% of the original value (which means 10\% nonuniformity) is added to the corresponding part from left-top to left-bottom in clockwise. From Fig.~\ref{fig_Nonuniformity}, visual distortion appears when the nonuniformity is larger than $30\%$, but when the nonuniformity is less than $20\%$, the deviation is negligible. 
\par\indent
In the  experiments, the uniformity of the pad surface resistivity is limited to $10\%\thicksim50\%$ by current thick-film resistor technology. Fig.~\ref{fig_NonuniformitySR} shows the surface resistivity measurements of a resistive board. For this board, the nonuniformity is about $20\%$ except one pad in the lower-left corner of the board. 
\par\indent
To study the nonuniformity effect, a scan along x-direction by 5mm step is carried out and the result is shown in Fig.~\ref{fig_Linearity}. The nonlinearity between the setup position and the measured position is less than 1\% which reflects that the board with nonuniformity of the pad surface reisitivity of about $20\%$ has an acceptable influence on the detector performance.
\begin{center}
\includegraphics[width=8cm]{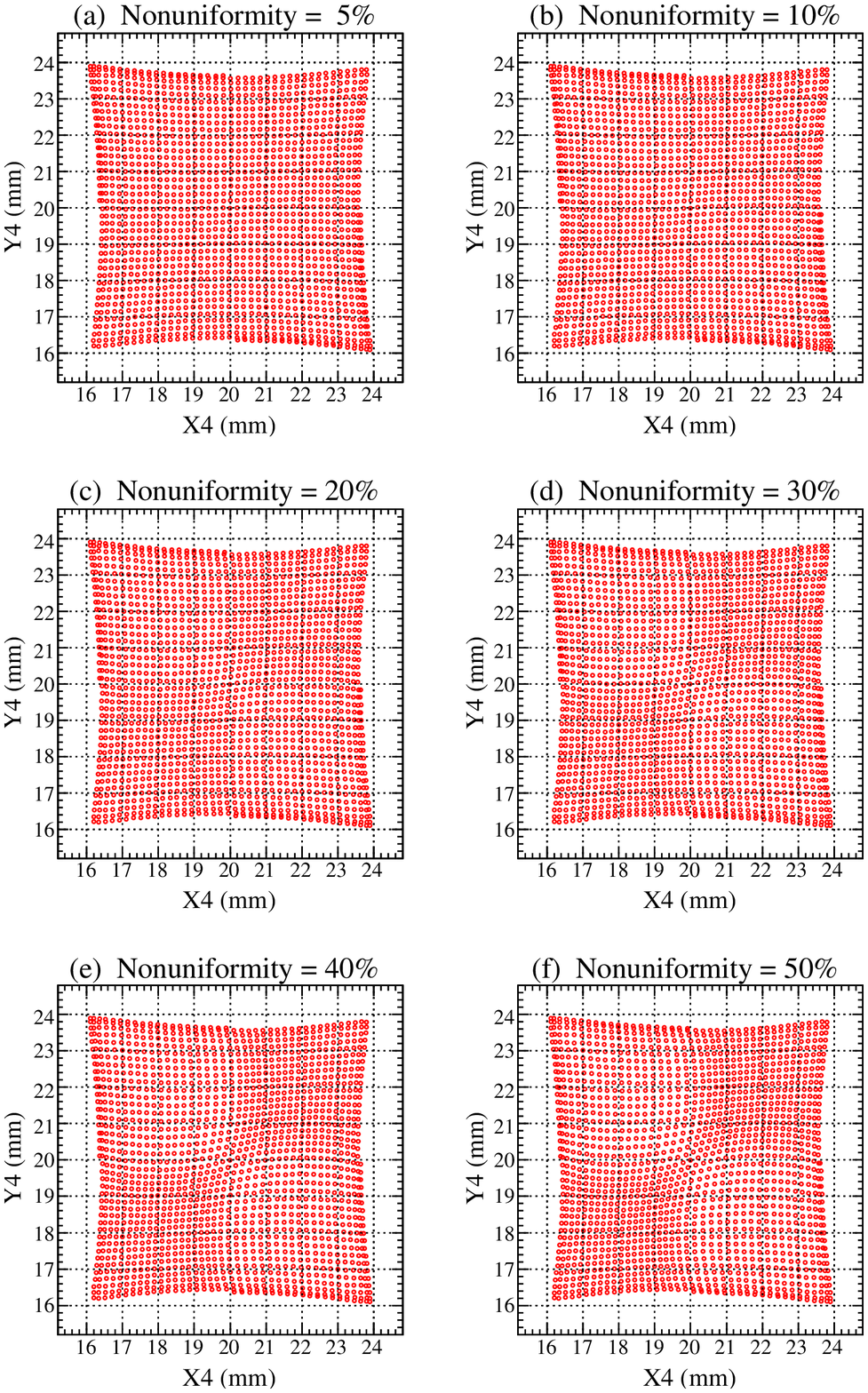}
\figcaption{\label{fig_Nonuniformity} Simulation results of nonuniformity of the pad surface resistivity effect on the imaging performance. The size of the pad and the strip is $8mm\times8mm$ and $8mm\times0.2mm$, respectively. The initial  $R_{{\square}P}$ is $100k\Omega/\square$ and $R_{{\square}L}$ is $1k\Omega/\square$. From (a) to (f), 5\%, 10\%, 20\%, 30\%, 40\%, 50\% deviations  are adding to original $R_{{\square}P}$, respectively.}
\end{center}
\begin{center}
\includegraphics[width=8cm]{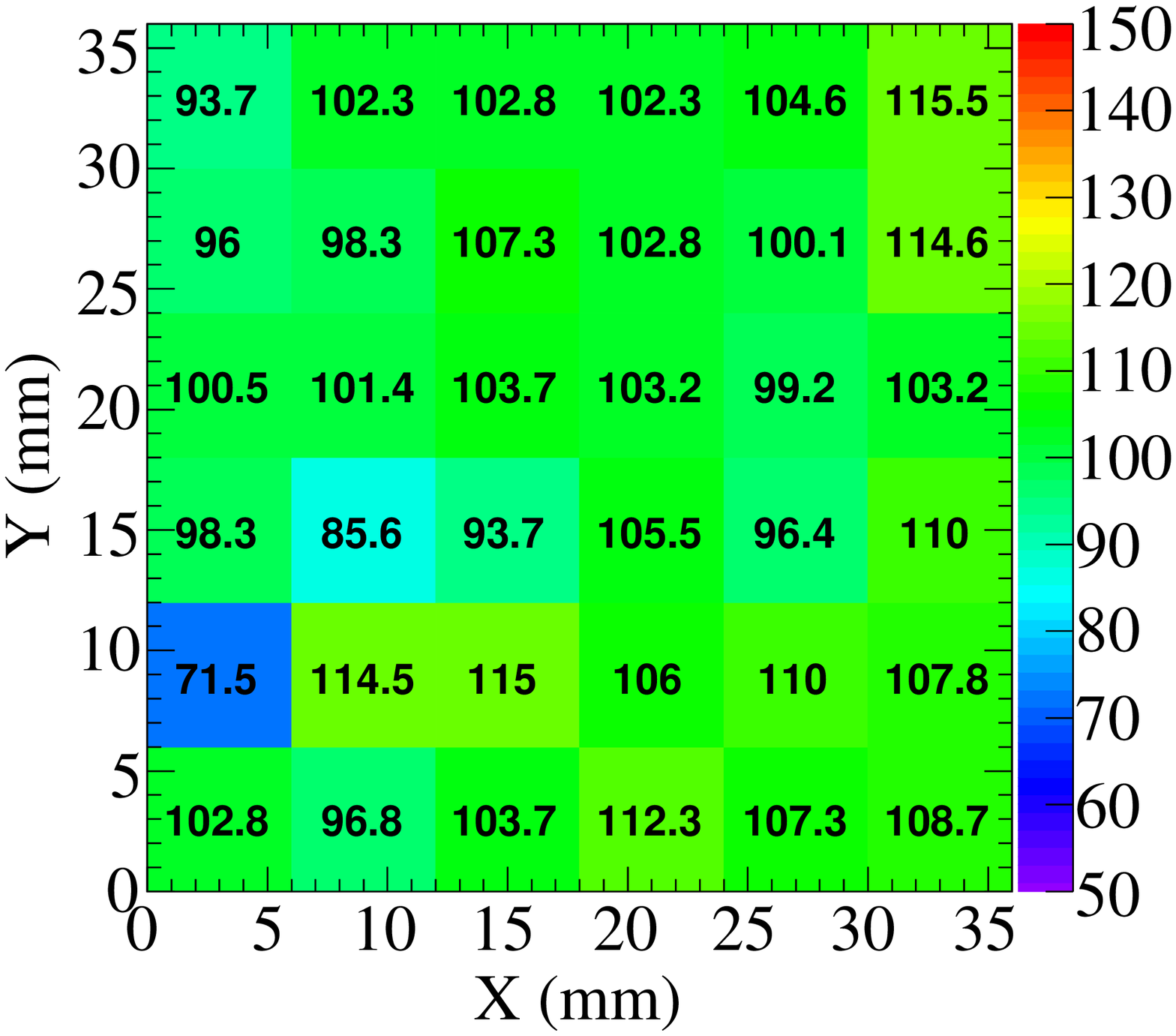}
\figcaption{\label{fig_NonuniformitySR} Measurements of the pad surface resistivity. Values in each pad are the measured pad surface resistivity in unit $k\Omega/\square$. For this board, the pad size is $6mm\times6mm$, the strip size is $6mm\times0.2mm$ and the design value of the pad surface resistivity is $100k\Omega/\square$.}
\end{center}
\begin{center}
\includegraphics[width=8cm]{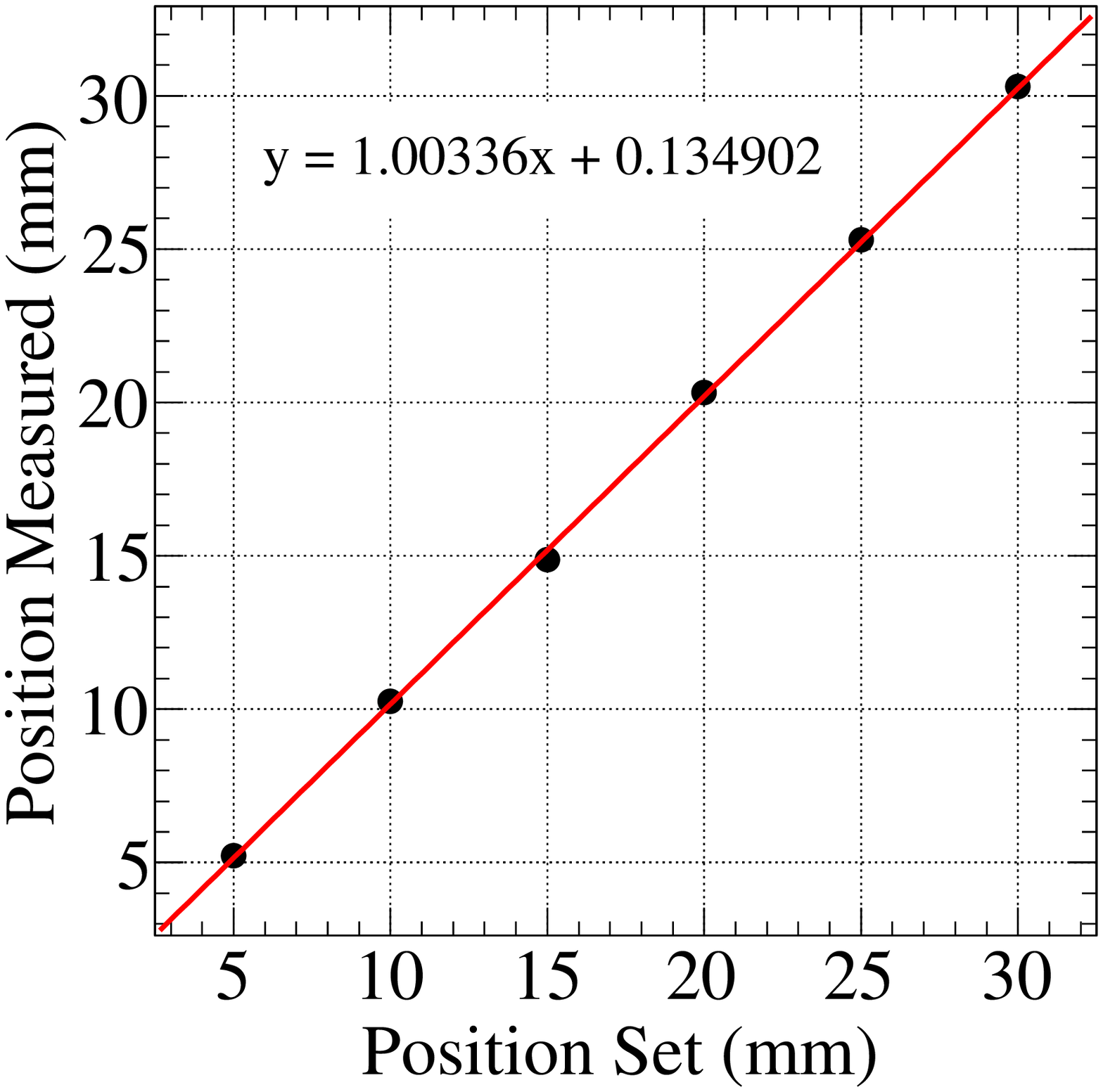}
\figcaption{\label{fig_Linearity} Position linearity scan by 5mm step. Results are getting from a $40{\mu}m$-slot by using X-ray machine for anode in Fig.~\ref{fig_NonuniformitySR}. }
\end{center}
\subsection{Imaging result}
Based on the results mentioned above, several optimized resistive anodes are fabricated and tested in a detector prototype with 49 electronics channels. Fig.~\ref{fig_RAGEM} shows the imaging result of a mask "AGE" with slot width of about $1mm$. For this resistive anode board, $R_{{\square}P}/(R_{{\square}L}{\cdot}N)\thickapprox6$, the pad size is $6mm\times6mm$, the strip size is $6mm\times0.2mm$ and the nonuniformity of the pad surface resistivity is less than $20\%$(Fig.~\ref{fig_NonuniformitySR}). The image is getting by X-ray machine which is about $40cm$ far away from the detector. The image quality is so good that the sharp borders of the letters are detailed. And the spatial resolution of the prototype is as good as $103.4{\mu}m$(Tbl.~\ref{tab_SR}).
\begin{center}
\includegraphics[width=8cm]{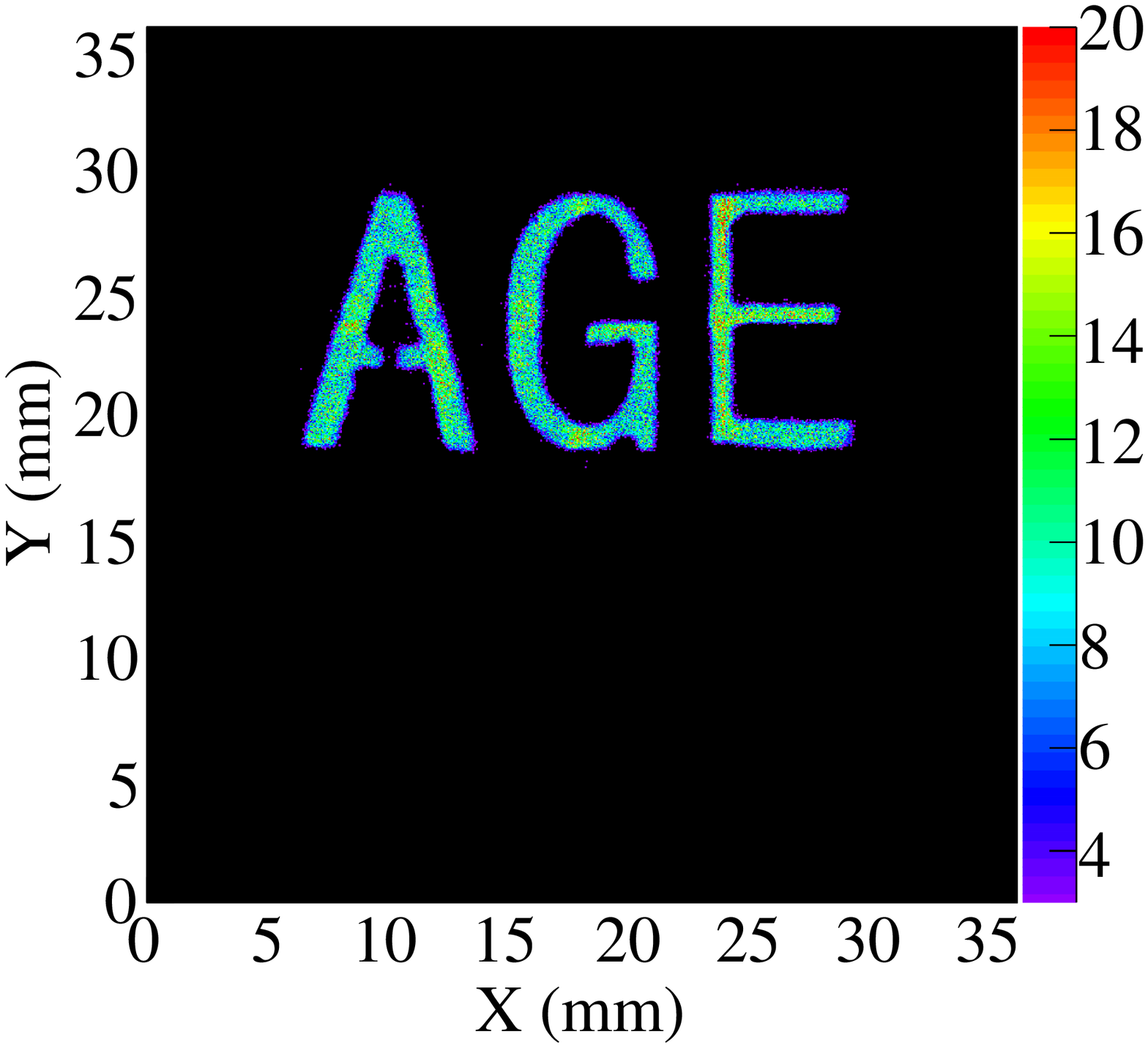}
\figcaption{\label{fig_RAGEM} Detector imaging of ``AGE''. The ``AGE'' pattern with a typical slot width of about $1mm$ and thickness of $12mm$ is printed by a 3D printer.}
\end{center}
\section{Conclusion}
The optimization of the resistive anode has been studied by both experiments and simulations. To obtain good detector performance, following rules should be considered for the resistive board:
\begin{itemize}
\item[-] Resistance ratio of the pad to the strip should be $\geqslant5$
\item[-] Nonuniformity of the pad surface resistivity had better be $\leqslant20\%$
\item[-] A smaller pad width leads to a smaller spatial resolution, when pad width is $6mm$, $\sigma$ can reach $105{\mu}m$.
\end{itemize}
\par\indent
Besides the factors mentioned above, from experiments, some other factors have been found contributing to the detector imaging as well. For example, narrower strips benefit the imaging, while current technology limits the strip width to about $0.2mm$. In general, the detector performance results from the comprehensive function of above factors.
\par\indent
The study helps to determine parameters of resistive readout boards and is enlightening for detectors using the similar readout method. With the spatial resolution about $110{\mu}m$, the study also shows the application prospect of the resistive anode readout method in MPGDs.
\par\indent
\acknowledgments
{
We wish to express our thanks to Dr. XIU Qing-Lei for the useful discussions and Dr. QI Hui-Rong for the electronics.
}
\end{multicols}

\vspace{-1mm}
\centerline{\rule{80mm}{0.1pt}}
\vspace{2mm}

\begin{multicols}{2}

\end{multicols}

\clearpage

\end{CJK*}

\begin{thebibliography}{90}
\vspace{3mm}
\newcommand{\etal}{et al}
\newcommand{\NIMA}{Nucl.\ Instrum.\ Meth.\ A}
\newcommand{\CPC}{Chin.\ Phys.\ C}
\bibitem{ref-1}
F.~Sauli \etal{}, 
\NIMA{}, 
{\bf 8}: 
386 
(1997)
\bibitem{ref-2}
D.~Abbaneo \etal{}, 
JINST, 
{\bf 8}: 
C12031 
(2013)
\bibitem{ref-3}
F. Thibaud \etal{}, 
JINST, 
{\bf 9}: 
C02005 
(2014)
\bibitem{ref-4}
A. Balla \etal{}, 
\NIMA{}, 
{\bf 628}: 
194---198 
(2011)
\bibitem{ref-5}
M. Lampton \etal{}, 
Review of Scientific Instruments, 
{\bf 50(9)}: 
1093---1097 
(1979)
\bibitem{ref-6}
T. Doke \etal{}, 
\NIMA{}, 
{\bf 261}: 
605---609 
(1987)
\bibitem{ref-7}
A.~Banu \etal{}, 
\NIMA{}, 
{\bf 593}: 
399 
(2008)
\bibitem{ref-8}
A.~Sarvestani \etal{}, 
\NIMA{}, 
{\bf 419}: 
444 
(1998)
\bibitem{ref-9}
DONG Ming-Yi \etal{}, 
\CPC{}, 
{\bf 37(2)}: 
026002 
(2013)
\bibitem{ref-10}
H.~Wagner \etal{}, 
\NIMA{}, 
{\bf 482}: 
334---346 
(2002)
\bibitem{ref-11}
XIU Qing-Lei \etal{}, 
\CPC{}, 
{\bf 37}: 
106002 
(2013)
\bibitem{ref-12}
http://garfieldpp.web.cern.ch/garfieldpp/documentation/UserGuide.pdf, received 17th May 2015
\bibitem{ref-13}
H.~Wagner \etal{}, 
\NIMA{}, 
{\bf 523}: 
287 
(2004)
\bibitem{ref-14}
WU Ling-Hui, 
\textit{Study of the offline calibration for the BES\uppercase\expandafter{\romannumeral3} drift chamber and the beam test of a prototype}, Ph.D. Thesis(Beijing:Institute of High Energy physics, CAS, 2007)(in Chinese)
\end{thebibliography}
\end{document}